\documentclass[11pt,onecolumn,groupedaddress,superscriptaddress,nofootinbib,showpacs
,preprintnumbers]{revtex4-1}
\usepackage{epsf}
\usepackage{graphicx}
\usepackage{amsfonts}
\usepackage{amssymb,ulem,color}
\usepackage{natbib}
\usepackage{dcolumn}
\usepackage{bm}
\usepackage{subeqnarray}
\newcommand{\bwt}{\begin{widetext}}
\newcommand{\ewt}{\end{widetext}}
\newcommand{\beq}{\begin{equation}}
\newcommand{\eeq}{\end{equation}}
\newcommand{\bea}{\begin{eqnarray}}
\newcommand{\eea}{\end{eqnarray}}
\begin{document}
\title{Constraints of the variation of fundamental couplings and sensitivity of the equation of state of dense matter}
\author{M. \'Angeles P\'erez-Garc\'{\i}a}
\email{mperezga@usal.es}
\affiliation{Departamento de F\'{\i}sica Fundamental and IUFFyM, Universidad de Salamanca, E-37008 Salamanca}
\author{C. J. A. P.  Martins}
\email{Carlos.Martins@astro.up.pt}
\affiliation{Centro de Astrof\'{\i}sica da Universidade do Porto, Rua das Estrelas, 4150-762 Porto, Portugal}

\date{\today}
\begin{abstract}
We discuss the coupled variations of the gravitational, strong and electroweak coupling constants and the current knowledge of the nuclear equation of state based on heavy ion collision experiments and neutron star mass-radius relationship.  In particular we focus in our description on phenomenological parameters, $R$, relating variations in the  quantum chromodynamics scale $\Lambda_{QCD}$ and  the fine structure constant $\alpha$,  and $S$, relating variation of $v$, the Higgs vacuum expectation value and the Yukawa couplings, $h$, in the quark sector. This parametrization is valid for any model where gauge coupling unification occurs at some (unspecified) high energy scale. From a physically motivated set of equations of state for dense matter we obtain the constrained parameter phase space $(R, S)$ in high density nuclear environments. This procedure is complementary to (although currently less powerful than) those used in low density conditions. For variations of $\Delta \alpha/\alpha=0.005$ we find that the obtained constrained parameter lies on a strip region in the $(R, S)$ plane that partially overlaps some of the allowed values of parameters derived from primordial abundances. This may be of interest in the context of unification scenarios where a dense phase of the universe may have existed at early times. 
\end{abstract}
\maketitle

\section{Introduction}

The properties of matter are the result of the interplay of fundamental interactions. Current knowledge points towards {\it new physics} beyond the standard paradigm of three fundamental forces \cite{pdb}, and suggests that unification should occur at high---but so far unknown---energies.
In that context, possible variations in the strong, electroweak and gravitational interactions are not independent of each other but, in a wide class of unification scenarios they are coupled.
In this sense  a variety of limits on the variation of the fine structure constant, $\alpha$, and the ratio of quark masses to the quantum chromodynamics (QCD) scale, $m_q/\Lambda_{QCD}$, have been established. In particular for the latter one there are  recent measurements from consideration of Big Bang nucleosynthesis (BBN), quasar absorption spectra, and the Sm isotopes in the Oklo natural nuclear reactor which was active about 1.8 billion years ago (for a review see e.g. \cite{uzan}). 

In order to model possibly coupled variations of fundamental constants in a framework valid for any model where gauge coupling unification occurs at some (unspecified) high energy scale it is assumed that the electroweak scale is derived by dimensional transmutation and that the varying couplings are due to some dilaton-type scalar field.  In this way a set of phenomenological parameters $(R, S)$ \cite{coc} relating Yukawa  couplings, $h$, Higgs vacuum expectation value, $v$, QCD mass scale, $\Lambda_{QCD}$, and fine structure constant, $\alpha$ is usually assumed. More in detail, $S$ is defined as $\Delta v/v=S\Delta h/h$ and since there is some model dependence a safe assumption is allowing $S$ to variate. Typically $S\sim\Delta_t$ where $\Delta_t$ is the sensitivity  from the top quark and defined as $\Delta_t={\partial\,\rm  ln \,m_W}/{\partial\, \rm ln \,a_i}$ , where $m_W$ is the mass of the W boson and $a_i$ are the input parameters of the supersymmetric model following the Barbieri-Giudice measure for an observable \cite{barb, ellis1}. Assuming a value for the Higgs vacuum expectation value of $v=M_P\,\rm exp{(-8\pi^2c/h^2_t)}$ \cite{vev}, where c and $h_t$ central values can be considered to be of order unity $c\simeq h_0 \simeq 1$ and $M_P$ is the Planck mass, and variating this expression it results that $\Delta v/v\simeq 160 \Delta h/h$ which will be taken as a reference for the canonical choice $S\sim 160$ from now on. Considering supersymmetric models with more massive particles may enlarge $S$ up to $\approx 500$. As for the $R$ parameter in some  grand unification models it is predicted that $\Lambda_{QCD}$, is modified as $\delta (m/\Lambda_{QCD})/(m/ \Lambda_{QCD}) \sim R\,\delta \alpha/\alpha$ where $m$ is the quark or electron mass in these  models. This result is strongly model dependent and it could yield $R\sim 36$ for some models \cite{fla} or even negative sign values as argued in \cite{dent3}. However, the large coefficients in these
expressions are generic for grand unification models and therefore signal that possible variation of masses and the strong interaction scale may be easier to detect than the variation of the fine structure constant $\alpha$.

Previous works \cite{coc} \cite{flambaum} focused on the low baryonic mass density enviroment of BBN  where $\rho_B\approx 10^{-5} \, \rm g/cm^3$ and physical quantities of interest such as the nucleon mass difference in vacuum $Q=m_n-m_p$, the neutron lifetime and the binding energy of deuterium. For example, Coc et al. \cite{coc} found that assuming a dilaton model and coupled variations one could accommodate the observational data of primordial abundances of light nuclei with an scenario of coupled variations for several parameter choices, including $\Delta h/h=1.5\, \,10^{-5}$, $R=16$,  $S=240$ and $\Delta h/h=2.5 \,10^{-5}$, $R=45$, $S=240$; in both cases $\Delta \alpha/\alpha=2 \Delta h/h$. 

In this work we use the complementarity provided by the extreme conditions of matter described by the nuclear equation of state (EoS) in heavy ion (HI) and neutron star (NS) physics. Even if the precision of inferred nuclear observables differs from the BBN nuclear abundance data or other techniques \cite{other}, one can assess the constraining power of these high density environments and scan the ($\alpha, R, S$) parameter space. A subsequent goal (which is beyond the scope of this paper) will be to carefully study particular implications for unification scenarios.

In an astrophysical scenario the EoS of nuclear matter is crucial for describing NS structure and, in particular, to obtain its mass-radius relationship. Conversely, kinematical observables, photospheric radius expansion or thermal emission measurements for NSs allow to infer compact masses and radii that constrain the possible values for central densities and the EoS itself. Typical NSs have on average a mass $M_{NS} \approx 1.4 M_{\odot}$ and radius $R_{NS}\approx 12 \,\rm km$. Roughly, their structure can be described as an external crust where matter has densities below saturation density, $n_0\approx 0.145 \, {\rm fm}^{-3}$ or, correspondingly, in mass density $\rho_0 \approx 2\,10^{14}\rm \,g/cm^{3}$, and an internal core where densities are larger and so far unknown. These objects are born in the aftermath of a supernova event with birth energies $\Delta E \simeq E_{grav} \approx \frac{3GM^2_{NS}}{5R_{NS}} \approx 10^{53}$ erg released when a progenitor exhausts the core reactions fueling the star and it can no longer support its own gravity. Some of them may have extreme external magnetic fields (of the order $B\approx 10^{9}-10^{15}$ G) and emit regular radiation pulses with periods $T\approx 10^{-3}-10^{1}$ s, being known as {\it pulsars}. The interior structure of NSs in the spherically symmetric static approximation is obtained by solving the Tolman (TOV) equations \cite{tov}, given an EoS $P=P(\epsilon, T)$ relating pressure, $P$, energy density, $\epsilon$, and temperature $T$.

This same EoS is also relevant for understanding the properties of nuclei and yields of HI collisions (at higher temperatures) experimentally accessible on Earth. Both sides of low and high densities allow to partially test the phase space of matter in the density-temperature plane.

In this work we consider the physics of matter under the extreme conditions that may resemble somewhat those of the early universe in the nuclear dense phase, by studying how the EoS can be used to constrain simultaneous coupled variations of various dimensionless fundamental couplings. Such a possibility, in particular a {\it time variation} of the fine structure constant, steams from observations of quasar absorption systems and a fit of positions of absorption lines \cite{webb} suggesting a smaller value in the past. Additionally, there are some indications that current data from Keck and VLT could be reinterpreted  as a possible {\it spatial variation} or gradient in the fine structure constant $\alpha$ \cite{alphaexp,berengut}. In this sense some of the parameters in models already describing coupled variations are obtained in the context of primordial abundances in BBN low conditions, these probe different conditions that for the high density interiors of NSs and HI collisions. As we explain in this work the possible variation of masses of different population species under beta equilibrium, namely nucleons and electrons, can be modified by the non vanishing values of a variation in $\alpha$. The in-medium effects can be parametrized by a meson-field model of strong interaction, due to the fact that high density environment is now considered, and typically one has to consider "dressed" or effective masses. In this way gravitational, weak and strong interactions are considered into the picture at high density conditions  at the phenomenological level. The modelization used has to deal with a number of uncertainties but the conclusions remain generic otherwise.

In section \ref{model} we introduce the phenomenological model used in this work to describe the set of EoSs for nuclear data and the generalized formalism of in-medium variations of quantities of interest in the high density conditions. In section \ref{results} we analyze the constraints of the $(R,S)$ parameter phase space to matter conditions as described by the set of EoSs used in this work. We apply our constraining procedure and discuss the agreement or overlap of some of the obtained values of $(R, S)$ parameters to those previously quoted in the literature. Finally, in section \ref{conclude}, we summarize and give some conclusions.

\section{Varying couplings}
\label{model}

Previous works trying to assess the importance of varying fundamental constants in the stellar structure \cite{adams,vieira} have considered a polytropic EoS $P=K\rho^{\Gamma}$, where $\rho$ is the mass density, $\Gamma=1+{1}/{n}$ and $n$ is the polytropic index. Matter at high densities can be usually  considered as a degenerate system where $T\approx 0$, since Fermi energies $E_{F\,i}$ of the degenerate $ith$-particle species are much higher than thermal energies  $E_{F\,i}/k_B T \gg 1$. The interplay of the constituent particle interactions is in-built in the EoS and further influences other observables as in HI collisions \cite{dani} or in a compact object mass and radius configuration \cite{steiner}. Typical central densities for these systems are not accurately known but for purely hadronic stars it is assumed \cite{zhang,lattimer,demorest} that they could reach up to $\approx 5 n_0$. About the composition, it has been hypothesized that more exotic objects composed of strongly interacting nucleonic, hyperonic or deconfined quark matter at their central regions \cite{perez10} may also exist. 

There are recent indications, coming from spectroscopic measurements along the line of sight of quasars, of space and time variations of $\alpha$ \cite{alphaexp}. These time variations are at the parts-per-million level. However, one must realize they apply to very low density environments. The only plausibly realistic way to explain such spatial variations would be with a chameleon-type field, and in that case the variations will also be environment-dependent,--they will depend on the local density \cite{book, barrow}. Therefore, the values of the couplings at high densities attained in the center of NSs or in HI collisions can differ from those on Earth or on low-density environments (as probed by quasar absorption systems or the cosmic microwave background (CMB)). In other words, the parts-per-million results of \cite{alphaexp} need not apply to the high density contrasts in NSs, and in what follows we will consider variations $\Delta \alpha/\alpha\approx \pm 10^{-3}$. Those can be accomodated by the CMB data in an analysis performed by Menegoni et al. \cite{cmb} who  obtained a result $-0.013 < \Delta \alpha/\alpha < 0.015$ at 95$\%$ C.L. from WMAP 5-year date combined with ACBAR, QUAD and BICEP experiments data or in the framework of some other models \cite{yoko}\cite{berengut}. Again we stress that, as pointed out in \cite{berengut,book}, there is no inconsistency with a value of $\alpha$ differing in such a proportion as those compared to quasar absorption since they could  probe {\it different conditions}.

We now address the variation of fundamental constants in terms of density dependent couplings. To parametrize varying couplings we will use the fact that when a dimensionless coupling such as the fine-structure constant is changed by a small amount $\alpha=\alpha_0(1+\delta_{\alpha})$, changes in other quantities can be related to it through coefficients ${\Delta X}/{X}=k_X {\Delta \alpha}/{\alpha}$ for $X=X_0(1+k_X\delta_\alpha)$. This is the case in unification scenarios shown in \cite{coc}, for which these changes can be phenomenologically described by two parameters: $R$, relating $\Delta \Lambda_{QCD}$ and $\Delta \alpha$,  and S relating $\Delta v$ and $\Delta h$ (the Yukawa couplings, all assumed to be the same) that we will discuss below. 

Matter in a NS is subject to gravitational, electroweak and strong interaction. Therefore, we will consider coupled variations in the fine structure constant $\alpha={e^2}/{\hbar c}$ and, additionally, allow for variation in particle masses on the lepton and hadron sectors. We define a set of dimensionless quantities to work with, namely the proton to electron mass ratio $\mu={m_p}/{m_e}$, and for physically meaningful changes in the gravitational constant $G$ we consider the ratio $\alpha_G ={Gm^2_i}/{\hbar c}={m^2_i}/{M^2_{P}}$ for the $ith$-type particle species. From the work of \cite{coc} the variations in ${\mu}$ can be written as ${\Delta \mu}/{\mu}=[0.8R-0.3 (1+S)]{\Delta \alpha}/{\alpha}$. 

On physical grounds both $R$ and $S$ may be expected to be positive, however there are some simplified phenomenological descriptions \cite{yoko} where one has, for example, $S=-1$ and $R=109$ (although they have some drawbacks as compared with quasar data). Let us emphasize that there is no experimental data which unambiguously shows what the correct unification model is like (or indeed, if one wants to be skeptical, whether high-energy unification happens at all). In this work we have adopted a {\it phenomenological} approach, treating  $R$ and $S$ as free parameters, and ask at what extent they are constrained by current knowledge of nuclear EoS physics, and how robust are these constraints. A list of models, representative of several possible unification scenarios, where $R$ can have positive or negative values of several hundreds, are discussed in \cite{coc, dent, dent2}. As mentioned before, in the case of $S$ its value is related to the neutralino mass and may be as large as $S\approx 500$ as pointed in \cite{ellis}. Here we conservatively assume that $R$, $S$ can take values, either positive or negative between the maximum absolute value, $R, S \in [-500, 500]$. As most observables of interest will depend on a linear combination of R and S we phenomenologically assume the same range of variation for both, as this will allow us to more easily identify possible degeneracy directions.

Despite some earlier results, there is currently no similar evidence for varying $\mu$ from a detailed analysis \cite{king, thompson}. If these results are correct they indicate that $\Delta\mu/\mu\ll\Delta\alpha/\alpha$ and therefore (in our class of models) $R\sim3(1+S)/8$, which could be used to eliminate one parameter from the analysis; however we will proceed without this constraint.

Let us notice that there is a vast number of model EoSs in the literature \cite{eos}. In this work we will firstly describe highly degenerate charge-neutral isospin asymmetric relativistic matter by a non-linear Walecka model (NLWM) \cite{bb} including baryons (neutrons and protons), mesons (Lorentz scalar, vector and iso-vector, $\sigma$, $\omega$ and $\vec \rho$ respectively) and leptons (electrons). We choose a parametrization, TM1 \cite{rmf}, which considers isospin symmetry since we have verified that more advanced parameterizations as PK1 \cite{pk1} yield same trend  results. Since a full comprehensive treatment is not affordable due to the large amount or current treatments of these systems we have accounted for the spread of the nuclear EoS by using a phenomenological approach. In section \ref{results} we define a pressure spread parameter, $\delta$ and allow it to vary from $\delta=0$ to $\delta=0.5$ to partially size the constraining power of this kind of description based 
on effective field theories. Let us point out that the main goal of each EoS is to describe the nuclear and astrophysical phenomenology currently known and there is no consensus about the actual degrees of freedom that should be present. Conservatively, we consider the NLWM a suitable EoS since it has proven to give a reasonable description of nuclear phenomenology but we keep in mind that the interior of NSs is largely unknown.
In \cite{coc} they use a non-relativistic potential model where the binding energy of deuterium is based  on the nucleon and $\sigma$ and $\omega$ meson mass. We introduce additionally the $\rho$ meson in our treatment since our system is largely neutron rich and therefore non isospin symmetric.

The variation in the meson-nucleon field coupling in the NLWM, $g_i$, $(i=\sigma,\omega,\rho)$ can be parametrized as $g^2_i=g^2_{i0}(m^2_i/M^2_{P}$), so that $ {\Delta g_i}/{g_i}= {\Delta m_i}/{m_i}$. The relative variations in the electron and proton masses are given by \cite{coc}, ${\Delta m_e}/{m_e}= 0.5(1+S)/{\Delta \alpha}/{\alpha}\,$, ${\Delta m_p}/{m_p}=  \left[ 0.8R+0.2(1+S)\right] {\Delta \alpha}/{\alpha}$, and since the nucleon mass difference parameter is $Q=m_n-m_p$ and variations are ${\Delta Q}/{Q}=(0.1+0.7S-0.6R) {\Delta \alpha}/{\alpha}$ we find that for neutrons ${\Delta m_n}/{m_n} = \left[ (0.1+0.7S-0.6R)+({m_p}/{m_n})(0.1-0.5S+1.4R) \right] {\Delta \alpha}/{\alpha}$. Note that in the isospin-symmetric case we recover $\Delta m_n/m_n=\Delta m_p/m_p$; moreover, requiring that the mass difference still vanishes as the couplings vary leads to the consistency condition $6R=7S+1$.

In the dense medium, the interacting Fermi systems of baryons have effective masses $m^*_p=m_p-g_{\sigma} \sigma, m^*_n=m_n-g_{\sigma} \sigma$ and the relative mass variation can be obtained for protons as, 
\begin{widetext}
\begin{equation}
 \frac{\Delta m^*_p}{m^*_p}=  \left[ \left(0.8R+0.2(1+S)\right)\frac{\Delta \alpha}{\alpha}- \frac{\Delta g_{\sigma}}{g_{\sigma}}\right]\frac{m_p}{m^*_p}+ \frac{\Delta g_{\sigma}}{g_{\sigma}},
\end{equation}
\end{widetext}
and in an analogous way for neutrons.

To include the meson mass variation we consider the Feynman-Hellmann theorem \cite{fh} where the expected value of the quark condensate (in the light sector u, d, s) for the proton state is obtained as $\langle p|{\bar q}q|p\rangle={\partial m_p}/{\partial m_q}$. Using a description of the $\sigma$ meson as a $SU(3)$ singlet \cite{flambaum} the meson mass variation can be related to the $s$-quark mass, $m_s$, as ${\partial m_{\sigma}}/{\partial m_s}=\langle\sigma|{\bar q}q|\sigma \rangle=2/3$. Taking values for the quark masses (using $\hbar=c=1$) $m_u\approx 4$ MeV, $m_d\approx 6$ MeV, $(m_u\approx m_d\approx m_q)$,  $m_s\approx 104$ MeV  we obtain, accordingly, for the $\sigma$ meson:
\begin{equation}
 \frac{\Delta m_{\sigma}}{m_{\sigma}}\approx \frac{2}{3}\frac{ m_s}{m_{\sigma}} \frac{\Delta m_s}{m_s}+\frac{2}{3} \frac{(m_u+m_d)}{m_{\sigma}} \frac{\Delta m_q}{m_q},
\end{equation}
for the $\omega$ meson, 
\begin{equation}
\frac{\Delta m_{\omega}}{m_{\omega}}\approx \frac{m_s}{m_{\omega}} \frac{\Delta m_s}{m_s},
\end{equation}
and for the $\rho$ meson 
\begin{equation}
\frac{\Delta m_{\rho}}{m_{\rho}}\approx \frac{m_s}{m_{\rho}} \frac{\Delta m_s}{m_s}.
\end{equation}
Using the assumption that all relative variations of the Yukawa couplings are similar \cite{coc} we have ${\Delta m_{s,q}}/{m_{s,q}}={0.5(1+S)}{\Delta {\alpha}}/{{\alpha}}$. Assuming small variations of the dimensionless couplings in this dense system, we now describe the set of equations to solve for a charge neutral spin-saturated beta-equilibrated system given an input baryonic particle number density, $n$. The self-consistent relativistic mean fields values $\sigma=\langle\sigma\rangle$, $\omega^0=\langle\omega^0\rangle$, $\rho^0=\langle \rho^0\rangle$ are obtained from the set $\sigma= {g_{\sigma}(1+{\Delta g_{\sigma}}/{g_{\sigma}})}n_s/{ {m'}^2_{\sigma}}\,$, ${\omega^0}= {g_{\omega}(1+{\Delta g_{\omega}}/{g_{\omega}}) n}/{{m'}^2_\omega}$, $m^{2}_{\rho} \rho^{0} = g_{\rho}\left(n_{p}-n_{n}\right)/2$, $\mu_{n}= \mu_{p}+\mu_{e}$. Particle number densities of different involved species are $n_i=\{n_p, n_n, n_e\}$ and each one is given in terms of the Fermi momentum, $k_{Fi}$ as $n_i={k^3_{Fi}}/{3 \pi^2}$. Conserved baryonic number densities and electrical charge neutrality imply additionally $n=n_p+n_n$, $n_p=n_e$. The effective meson masses, ${m'}_{\sigma}$, ${m'}_{\omega}$, include non-linear self-interaction terms in the NLWM \cite{bb}.

The explicit expressions for the chemical potentials are given by $\mu_p=E^p_F+g_{\omega} \omega^{0}+g_{\rho}\rho^{0}/2$, $\mu_{n}=E^n_F+g_{\omega} \omega^{0}-g_{\rho}\rho^{0}/2$, $\mu_e=E^e_F$ where  $E^{p}_{F}$, $E^{n}_{F}$ and $E^e_F$ are, respectively,  the proton, neutron and electron Fermi energies given by $E^{i}_F=\sqrt{k^2_{Fi}+m^{*2}_i (1+2\frac{\Delta m^*_i}{m^*_i})}$, and $m^{*}_e=m_e$.  Finally, the scalar density is given by,
\begin{equation}
n_s=\frac{1}{\pi^2}\sum_{i=p,n} \int_0^{k_{Fi}} \frac{m^{*}_i \left(1+\frac{\Delta m^{*}_i}{m^{*}_i}\right) k^2\,dk}{\sqrt{k^2+m^{*2}_i (1+2\frac{\Delta m^*_i}{m^*_i})}}.
\end{equation}

If the solution exists, it is given by values solving the self-consistent set of non-linear equations for the NLWM EoS $P=P(\epsilon, T=0)$. In this way the pressure is given by \cite{bb},
\begin{widetext}
\begin{equation}
P=\frac{1}{3\pi^2}\sum_{i=p,n} \int_0^{k_{Fi}} \frac{k^4\,dk}{\sqrt{k^2+m^{*2}_i (1+2\frac{\Delta m^*_i}{m^*_i})}}
+ \frac{1}{3\pi^2}\sum_{i=e} \int_0^{k_{Fi}} \frac{k^4\,dk}{\sqrt{k^2+m^{2}_i (1+2\frac{\Delta m_i}{m_i})}}+P_{m},
\end{equation}
where the mesonic field pressure is given in terms of the non-linear couplings $\kappa$, $\lambda$ and $\xi$ as,
\begin{equation}
P_{m}=\frac{-1}{2}m^2_{\sigma}(1+2\frac{\Delta m_\sigma}{m_\sigma})\sigma^2+\frac{1}{2} m^2_{\omega}(1+2\frac{\Delta m_\omega}{m_\omega})(\omega^0)^2\,
+ \frac{1}{2} m^2_{\rho}(1+2\frac{\Delta m_\rho}{m_\rho})(\rho^{0})^2-\frac{1}{3!}\kappa \sigma^{3}
 - \frac{1}{4!}\lambda \sigma^{4}+\frac{1}{4!}\xi g^4_{\omega}(\omega^{0})^4,
\end{equation}
and the energy density is,
\begin{equation}
\epsilon=\frac{1}{\pi^2}\sum_{i=p,n} \int_0^{k_{Fi}} {k^2\,dk}{\sqrt{k^2+m^{*2}_i (1+2\frac{\Delta m^*_i}{m^*_i})}}
+\frac{1}{\pi^2}\sum_{i=e} \int_0^{k_{Fi}} {k^2\,dk}{\sqrt{k^2+m^{2}_i (1+2\frac{\Delta m_i}{m_i})}} +\epsilon_m\,,
\end{equation}
where the mesonic contribution is $\epsilon_m = -P_m$.
\end{widetext}

\section{Results}
\label{results}
\begin{figure}[hbtp]
\begin{center}
\includegraphics [angle=-90,scale=0.75]{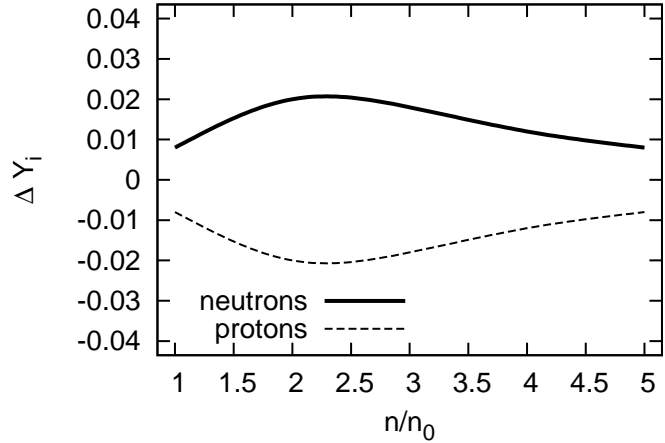}
\caption{Variation of particle population fractions with $\Delta \alpha/\alpha=+0.005$, R=20, S=160 as a function of baryonic number density.}
\label{Fig2}
\end{center}
\end{figure}

We now discuss the results obtained from exploring the ($\alpha, R, S$) space of the dense system. We note that our analysis differs from that of Coc et al. \cite{coc} where some unification models were studied, by choosing fixed values of $\alpha$, $R$ and $S$. Here our goal is to {\it scan} the phenomenological $(R,S)$ space for fixed choices of $\alpha$, in order to ascertain the feasibility of constraining this space using the current knowledge of the nuclear EoS.
A fine-structure constant variation of  $\Delta \alpha/\alpha=+0.005$ is considered. In Fig.\ref{Fig2} we show the variation of particle population for protons (dashed line) $\Delta Y_{p}$ or neutrons  $\Delta Y_{n}$ (solid line)  defined with respect to the unchanged value of $\alpha$, $\Delta Y_{i}=Y^{\Delta \alpha/\alpha}_i-Y^{\Delta \alpha/\alpha=0}_i$ $i=p,n$ as a function of particle number density.  We use $\Delta \alpha/\alpha=+0.005$  and a 'canonical' choice of parameters $R=20, S=160$ \cite{coc}. It can be seen that  the particle population changes in an appreciable way decreasing the number of protons as $\alpha$ is enhanced. In a similar fashion to \cite{berengut}  a change in $\alpha$ or quark masses translates into a change in BBN abundances.

\begin{figure}[hbtp]
\begin{center}
\includegraphics [angle=-90,scale=0.75]{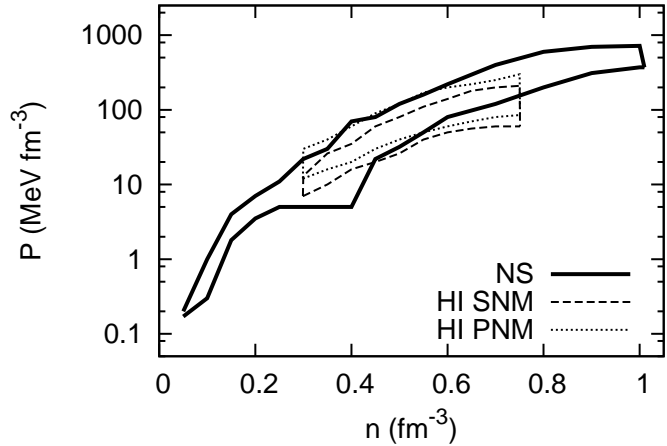}
\caption{Experimental nuclear EoS constraints in pressure from NS mass-radii measurements \cite{steiner} and HI collisions in the SNM and PNM case \cite{dani} as a function of baryonic number density.}
\label{Fig3}
\end{center}
\end{figure}

In order to illustrate the current experimental constraints on the nuclear EoS we plot in Fig. \ref{Fig3} the pressure versus the baryonic particle number density from a variety of input data. With solid line  we plot EoS data  from the work of Steiner et al. \cite{steiner} deduced from a selected group of NS mass and radius measurements. EoS data from HI collisions has also been plotted from the symmetric nuclear matter (SNM)  case (dashed line) and pure neutron matter (PNM) case (dotted line). Note that NS matter is in beta equilibrium and therefore none of the two ideal cases apply, but it is a system close to the PNM case. Both data  seem to suggest a soft EoS \cite{prak, sym}.

\begin{figure}
\begin{minipage}[b]{1.0 \linewidth} 
\centering
\includegraphics [angle=-90,scale=0.75]{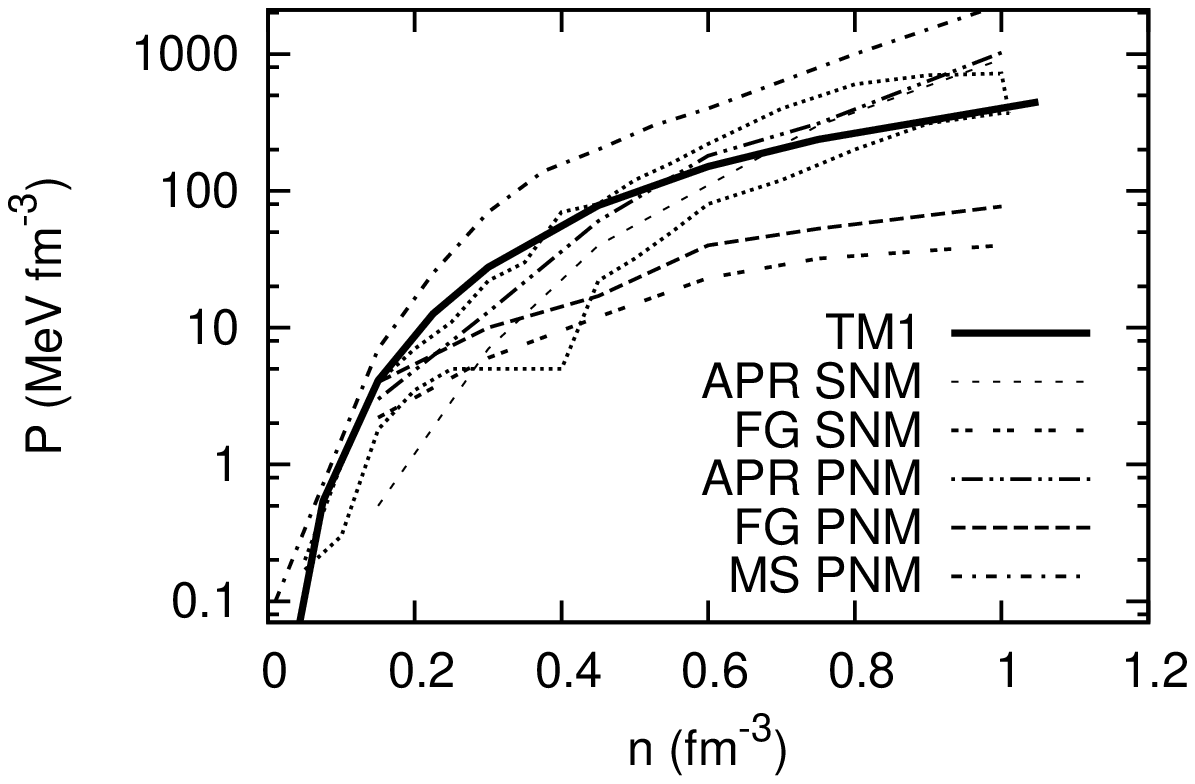}
\end{minipage}
\hspace{0.1cm} 
\begin{minipage}[b]{1.0 \linewidth}
\centering
\includegraphics  [angle=-90,scale=0.75]{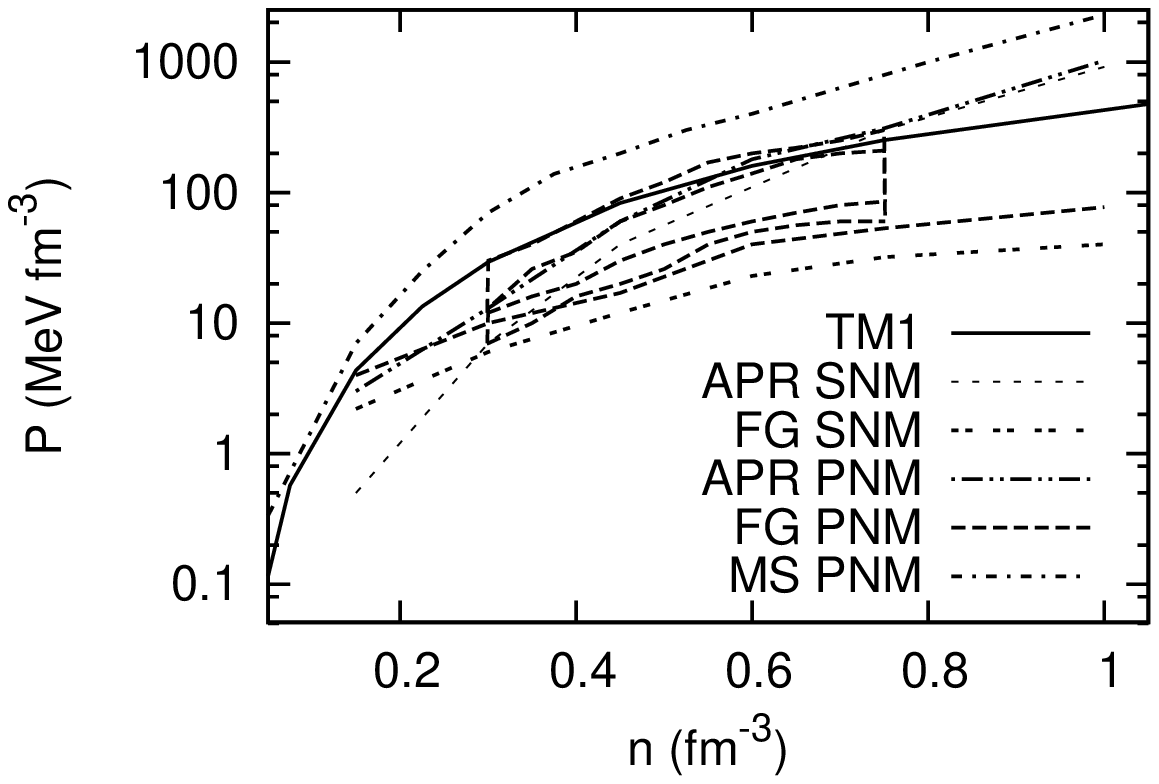}
\caption{Some frequently used nuclear EoS and NS constraints as deduced from \cite{steiner} (upper panel) and HI constraints as deduced from \cite{dani} (lower panel).}
\label{Fig5}
\end{minipage}
\end{figure}

There is a vast number of parameterizations of the nuclear EoS in the literature, including several treatments of the nuclear interaction. From those using the non-relativistic potential interaction models \cite{akmal} to the Fermi gas or the relativistic fields as Muller and Serot ($\zeta=0, \xi=0$) \cite{muller} or the Shen \cite{rmf} TM1 EoS. They all aim to describe a many-body system to give insight to the nuclear observables and the astrophysically deduced masses and radii of NSs.
In Fig. \ref{Fig5} (upper panel) we include a representation of some of the most representative nuclear EoS as obtained for the SNM and PNM cases and for beta equilibrium TM1 for the sake of comparison. We also plot the EoS NS constraints (dotted line). We see that most EoS do not describe the whole range of densities in complete  agreement. Some of them are not appropriate as e. g. the PNM EoS of Muller and Serot ($\zeta=0, \xi=0$), since it is too stiff or the Fermi gas EoS since it does not provide enough pressure at large densities.
We can see in the lower panel in Fig. \ref{Fig5} again a representation of the same set of EoS but using the HI constraints (dashed boxes) for the SNM and PNM cases. We see that again the TM1 EoS allows the representation the neutron rich side of those boxes, however some of the Muller and Fermi gas EoS do not describe the tendency of these regions as happened with NS constraints.

\begin{figure}[hbtp]
\begin{center}
\includegraphics [angle=-90,scale=0.75]{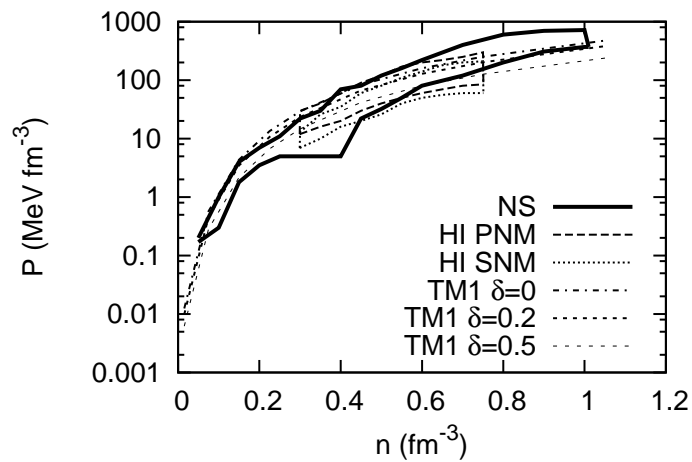}
\caption{Set of phenomenologically softened TM1 EoS $(\delta=0, 0.2, 0.5)$ versus EoS coming from HI and NS constraints.}
\label{Fig6}
\end{center}
\end{figure}

In order to partially consider the spread of the nuclear EoS we introduce a phenomenological parameter $\delta$ to parameterize the softness of the EoS and its effect is shown in Fig. \ref{Fig6}. In this way for the range $0\le \delta \le0.5$  the pressure is modified accordingly as  $P(1-\delta)$. Therefore $\delta=0$ corresponds to no modification of the pressure as given by the EoS considered and $\delta=0.5$ a $50\%$ softened EoS. Note that since our reference EoS describes better the stiff side of the constraints, we have $\delta \ge 0$. We consider the  phenomenological spread in the nuclear EoS and see that the unchanged TM1 EoS describes the neutron rich systems and as $\delta > 0$ more isospin symmetric systems can be accessed. With this procedure we partially size the combined effect of the uncertainty in the isospin content and the softness of the EoS and the robustness of our findings.

 Having gained a feeling for the effect of variations in $\alpha$, we now explore the $R, S$ parameter space. Interestingly, and although the effects are relatively mild for the 'canonical' choice of $R$ and $S$, we will now see that there are regions of parameter space where there is no possible solution to the set of equations showing, therefore, 'exclusion regions'.

\begin{figure}[hbtp]
\begin{center}
\includegraphics [angle=-90,scale=0.75]{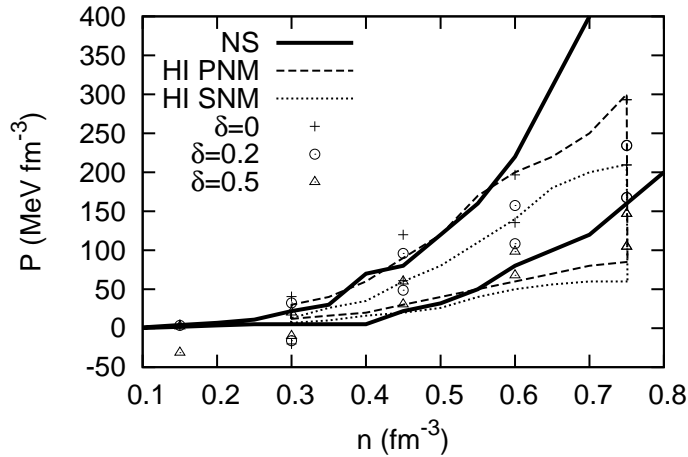}
\caption{Maximum and minimum values of pressure spread in the phenomenologically softened TM1 $(\delta=0, 0.2, 0.5)$ when $R$, $S$ are varied in the $[-500,500]$ interval for $\Delta \alpha/\alpha=0.005$ as compared to EoS from NS and HI constraints.}
\label{Fig7}
\end{center}
\end{figure}

In  Fig. \ref{Fig7}  we show for a value $\Delta \alpha/\alpha=0.005$ and $R, S$ in the interval $[-500, 500]$ using TM1 the maximum and minimum values for pressure as a function of density for selected values $n/n_0=1,2,3,4,5$ and for spreads $\delta=0, 0.2, 0.5$, crosses, circles and triangles respectively. The aim of this figure is to show the pressure spread and that not all unification scenarios, i.e. values of $R$ and $S$ are consistent {\it simultaneously} with existing EoS deduced from NS and HI datasets in the bounded areas. This is due to the sensitivity of the nuclear EoS to the strong effect on the effective particle masses in the dense medium from varying the couplings.

\begin{figure}[hbtp]
\begin{center}
\includegraphics [angle=-90,scale=0.75]{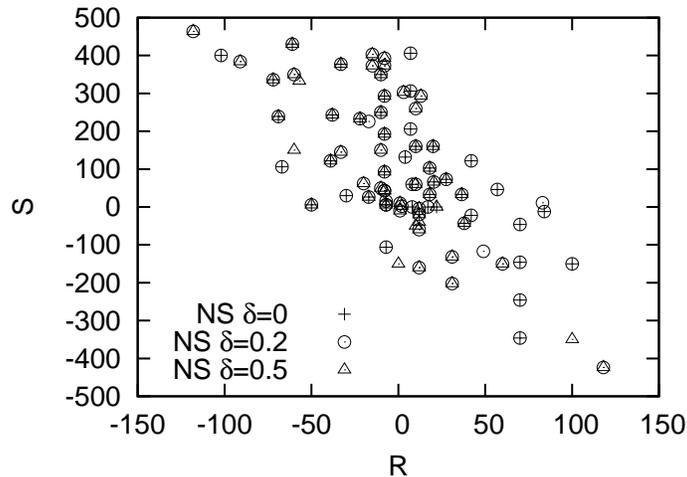}
\caption{$R,S$ phase space exclusion region.  Crosses, circles and triangles denote values of pressure compatible with NS constraints as results from applying a spread of $\delta=0, 0.2, 0.5$ for $\Delta \alpha/\alpha=0.5\%$.}
\label{Fig11}
\end{center}
\end{figure}

\begin{figure}
\begin{minipage}[b]{1.0 \linewidth} 
\centering
\includegraphics[scale=0.75, angle=-90]{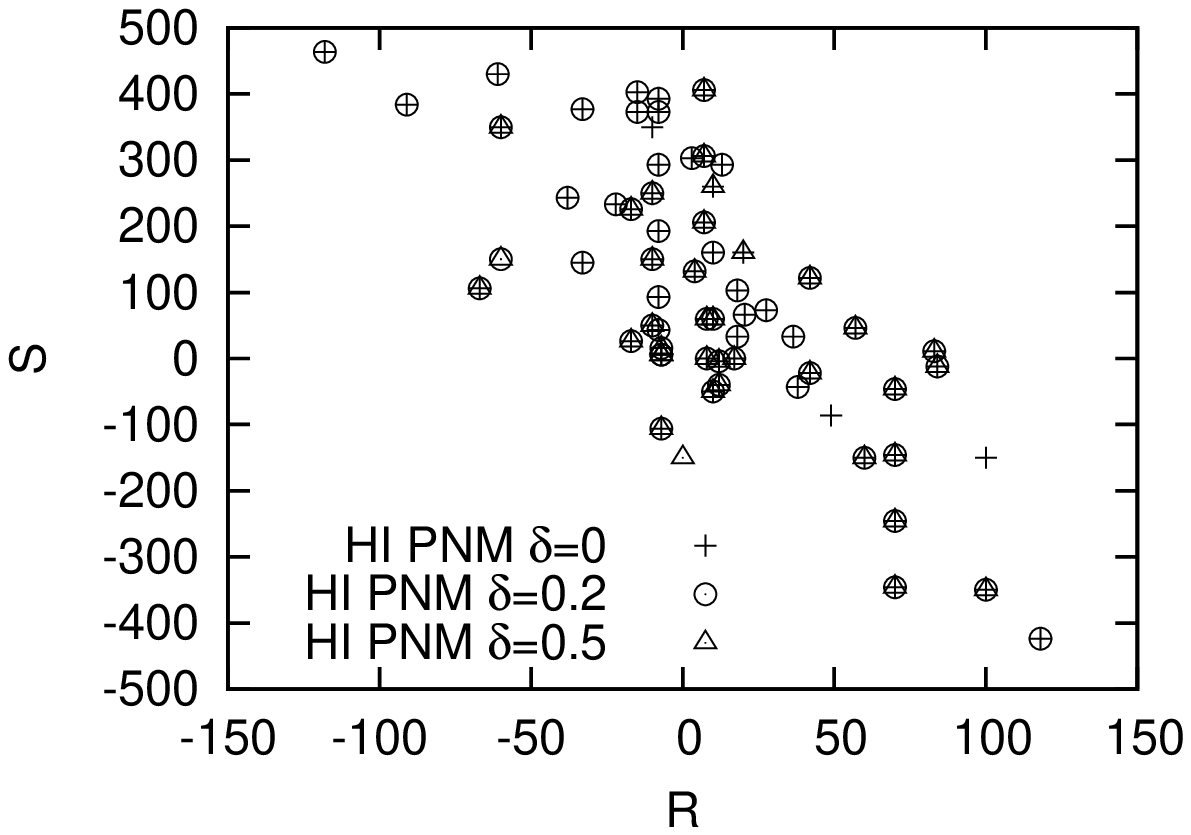}
\end{minipage}
\hspace{0.1cm} 
\begin{minipage}[b]{1.0 \linewidth}
\centering
\includegraphics[scale=0.75, angle=-90]{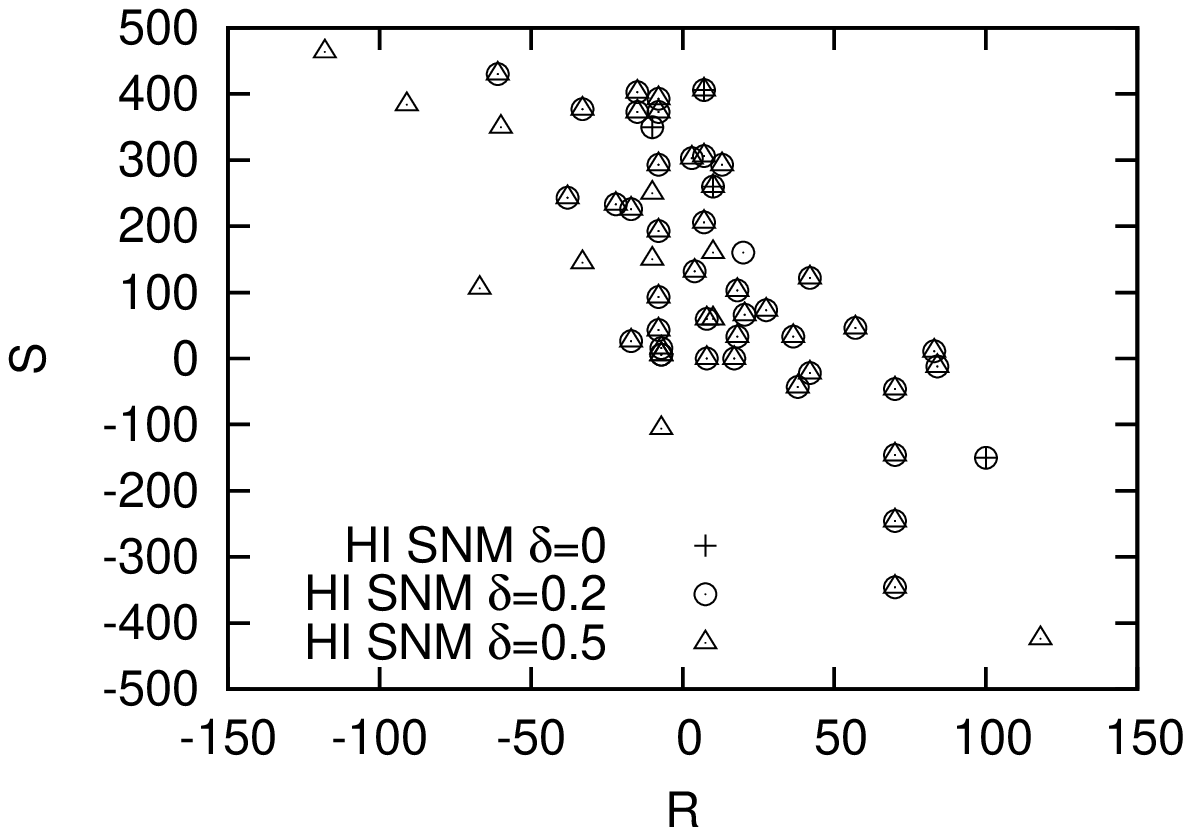}
\caption{Same as in Fig. \ref{Fig11} but for HI PNM (top panel) SNM (bottom panel) case for $\Delta \alpha/\alpha=0.5\%$. See details in the text.}
\label{Fig12}
\end{minipage}
\end{figure}

In Fig. \ref{Fig11} we plot with crosses, circles and triangles values of $R$ and $S$ yielding pressure values compatible with EoS from NS constraints using TM1 EoS with $\Delta \alpha/\alpha=0.005$ and  applying a spread of $\delta=0, 0.2, 0.5$ respectively. We see that, globally, the values compatible with EoS from NS constraints lie on a variable width strip. Without being fully comprehensive on the scan of the allowed $R$ and $S$ values we can see that equally stiff or soft EoS have constraining power in their phase space. Some of these values follow the same robust tendency when an extreme $\delta=0.5$ variation in the EoS is considered and they are superimposed.
However one must note that  the experimental uncertainty of these data is much larger that in the low density BBN case. Note for example that there are quantities that are especially sensitive to the parameter change as, for example, the $^7$Li abundance that is sensitive to a relative change of $50\%$ when $\alpha$ is changed $1\%$ \cite{berengut}. The same constraints are applied in the HI collision data show in Fig. \ref{Fig12}. We see that in either case the tendency is the same, showing robustness. The range of $R$ shown in both figures is somewhat lower than the originally prescribed due to exclusion of those larger values since they do not provide a valid self-consistent solution of the set of population equations, namely due to the contribution of effective masses of the particles.

Note that as the density increases the uncertainty of these measurements is larger and there is a trivial constraint on $R, S$ values since pressures must remain positive. We note that NS matter is mostly neutron matter but indeed not pure neutron nor symmetric matter (those cases constrained in \cite{dani} or at low density in e.g. \cite{habe}) but we use these constraints as an indication of the behavior. 

Finally following the analysis performed in \cite{coc} we find that both set of parameters $\Delta h/h=1.5\, \,10^{-5}$, $R=16$,  $S=240$ and $\Delta h/h=2.5 \,10^{-5}$, $R=45$, $S=240$  with $\Delta \alpha/\alpha=2 \Delta h/h$ are indeed  contained in our set of possible solutions. Using their assumed value $\Delta \alpha/\alpha\approx 10^{-5}$ would result in pressure deviations smaller than $1\%$, and therefore it would produce a negligible variation on the  EoS. Therefore the  constraining power  of the procedure presented in this work is meaningful for values of $\Delta \alpha/\alpha$ larger than this and currently present in some theoretical models of unification and data analysis \cite{cmb}.

As mentioned, we expect that some mild dependence on the EoS modelization may arise, in particular a more enriched description could somewhat enlarge or reduce those constrained regions  but would not produce a change of tendency in the constrained values since these descriptions use the same basic strong interaction, quark masses, fine structure and gravitational constant ingredients.

\section{Conclusions}
\label{conclude}
In this work we have performed a study of combined variations of the gravitational, strong and electroweak coupling constants based on the current constraints on the nuclear high density EoS. Using a phenomenological approach and assuming a variation of the fine structure constant  compatible, for example, with recent analysis of CMB data, $\Delta \alpha/\alpha\approx 10^{-3}$, we find that the valid solutions provided a variation of couplings in the studied model, lie on a variable width strip in the phase space of $R$,  relating variations in the  $\Lambda_{QCD}$ and  the fine structure constant $\alpha$,  and $S$, relating variation of $v$, the Higgs vacuum expectation value and the Yukawa couplings, $h$, in the quark sector. If we further restrict our results to positive R, S values on the first quadrant, based on more stringent physical assumptions for the most feasible grand unification models, we find that some of the parameters already obtained in the context of primordial abundances in BBN low conditions remain valid when considered in the high density conditions of NS interiors and HI collisions.
Although the experimental precision of nuclear observables in HI collision and NS physics is lower than in nuclear finite systems or primordial nuclear abundances, the extended system EoS depends crucially on in-medium density effects. Therefore we find that EoS from HI and NS data may constrain the variation of $\alpha$, $R$, $S$ parameters at densities beyond $n_0$ and provide additional non-trivial constraints on them.  

Our results show the potential for using the high density region of the density phase space of matter to partially constrain fundamental physics {\it in addition} to low density tests using the CMB. Future work to explore the relevant parameter space in more detail, and discuss how these tests can complement other constraints from stellar reactions is needed.

\section*{Acknowledgments}

We thank the FCT (Portugal)--MICINN (Spain) cooperation grant AIC10-D-000443 (Proc. 441.00 Espanha), FCT grant PTDC/FIS/111725/2009, COMPSTAR and MULTIDARK projects, FIS-2009-07238 and FIS2011-14759-E. The work of CJM is funded by a Ci\^encia2007 Research Contract, funded by FCT/MCTES (Portugal) and POPH/FSE (EC). 

\end{document}